\documentclass[twoside]{elsart}
\usepackage{epsfig}
\usepackage{array}
\usepackage{amssymb,amsbsy,amsmath}
\newcommand{\Hil}{\mathcal{H}}
\newcommand{\Na}{\mathbb{N}}
\begin{document}
%
%
%
\journal{PHYSICA A}
\begin{frontmatter}
\title{A solvable model of a one-dimensional quantum gas with pair interaction}

\author{H.-J.~Schmidt\thanksref{HJS}}
\author{ \and J.~Bartke}\\

\address{Universit\"at Osnabr\"uck, Fachbereich Physik \\
         Barbarastr. 7, D-49069 Osnabr\"uck}

\thanks[HJS]{email: hschmidt\char'100uos.de,\\
            WWW:~http://www.physik.uni-osnabrueck.de/makrosysteme}

\begin{abstract}

\noindent
We propose a solvable model of a one-dimensional harmonic oscillator
quantum gas of two sorts of particles,
fermions or bosons, which allows to describe the formation of pairs due to a suitable pair
interaction. These pairs we call ``pseudo-bosons" since the system can be approximated by an ideal
bose gas for low temperatures. We illustrate this fact by considering the specific heat and the
entropy function for $N=8$ pairs.
The model can also be evaluated in the thermodynamic limit if the harmonic oscillator potential
is suitable scaled.

\vspace{1ex}
\vspace{1ex}

\end{abstract}
\end{frontmatter}
\raggedbottom

\section{Introduction\label{I}}

It seems to be generally taken for granted that a compound of an even number of fermions behaves like a boson
since any state of $2nN$ fermions has the same permutational symmetry as a state of $N$ bosons.
Accordingly, the recent experiments
with trapped ultracold gases \cite{{AEM:S95}} \cite{DMA:PRL95} \cite{KeD:PRA96} \cite{BSH:PRL97}
are viewed as an achievement of Bose-Einstein condensation \cite{DGP:RMP99}.
However, a moment's thinking reveals that this cannot be true in an exact sense but at most approximately.
Consider an ideal gas of $N$ ``bosons", each ``boson" consisting of $2n$ fermions bound together by some attractive
force. In the high temperature limit the bounds between the fermions will be broken and the system behaves like
an ideal fermi gas at high temperatures. Its energy will thus be $E\sim 2 n N f_F k_B T$,
where $f_F$ is number of degrees
of freedom for a single fermion, e.~g.~$f_F=6$ for fermions sitting in a
$3$-dimensional harmonic oscillator potential which we assume now for sake of simplicity.
This would be consistent with the corresponding formula for a bose gas of $N$ particles if
the number of degrees of freedom for a single boson is assumed to be $f_B = 2 n f_F$. For intermediate temperatures
$T$ which are large with respect to the spacing of $1$-particle energies, say $k_B T\gg  \hbar \omega$,
but small with respect to the binding energy of the fermionic compounds, we would expect
a ``preliminary" classical limit of the energy, say  $E\sim 6 N  k_B T$, different from
$E\sim 12 n N k_B T$. Hence the energy function $E(T)$ of the considered gas cannot coincide with that
of an ideal bose gas at medium and high temperature.\\
More interesting with regard to Bose-Einstein condensation is the low temperature limit. $N$ non-interacting
bosons would occupy the same $1$-particle ground state for $T\rightarrow 0$, which is impossible for
$2nN$ fermions according to the Pauli principle. Hence we expect a non-bose behaviour for ``bosons" also for
low temperatures. In view of these problems we have scruples about identifying compounds of $2n$ fermions with
bosons and will rather call them ``pseudo-bosons".\\
To be sure, there will be a close resemblance between bosons and pseudo-bosons and the thermodynamic functions
of ideal gases of each sort should be approximately the same for some range of temperature,
not too low and not too high. To make this statement more precise is a challenge to the foundations of
quantum statistical mechanics, in particular, to demonstrate the possibility of pseudo-Bose-Einstein condensation.
We do not know of any attempt to tackle this problem, with the exception of an estimate of the critical density
of a pseudo-Bose-Einstein condensate where fermionic effects would occur \cite{RNPP:02}.\\
As a first step to treating the problem of pseudo-bosons we seek for simple models describing the formation
of pairs of fermions, being the simplest case of pseudo-bosons, which are solvable in the sense that the
partition function of the system can be explicitely calculated. For this purpose the model need not be a
realistic description of, say, rubidium atoms in a trap. It rather serves as an example to study the problem
of pseudo-bosons in principle.\\
In section \ref{S} (summary and outlook) we will come back to the problem of pseudo-bosons. Of course,
the model we are going to present
can be considered in its own sake without reference to the specific problem and could be compared with other
$1$-dimensional solvable models of quantum systems, see \cite{Mat:93}.
It turns out that the case of pairs of interacting bosons is completely analogous and can be treated
simultaneously, although the discussion of the results will be confined to the fermi case.
The model will be defined in section
\ref{D} and consists of two sorts of $N$ fermions (resp.~bosons) with equally spaced non-degenerate
$1$-particle energy levels. In this aspect it remotely resembles the Luttinger model
\cite{Lut:JMP63},\cite{ML:JMP65}, which
is, however, unitarily and hence thermodynamically equivalent to a system of non-interacting particles. In our
model the interaction is introduced in such a way that pairs of fermions of different sort with the same
quantum number gain a binding energy $V$. Hence the diagonalization of the total Hamiltonian is trivial.
The calculation of the partition function in section \ref{Z}, however, requires some combinatorics.
For $N\le 8$ we calculate the specific heat, using computer-algebraic means, and discuss its properties
(section \ref{C}). For larger $N$ one has to resort to the thermodynamic limit which is derived in section \ref{L}.
The convergence to this limit is shown to be rather rapid for $T>0$.

\section{Definition of the model\label{D}}

We will consider two sorts of particles with Hilbert spaces
$\Hil^{(1)}$ and $\Hil^{(2)}$
for $1$-particle states.
Let
$\left( \phi_n^{(1)} \right)_{n\in\Na}$
and
$\left( \phi_n^{(2)} \right)_{n\in\Na}$
be orthonormal bases in
$\Hil^{(1)}$ and $\Hil^{(2)}$.
Define the $1$-particle Hamiltonians by
\begin{equation}\label{D.1}
H^{(i)} \phi_n^{(i)} = n\hbar\omega \phi_n^{(i)},\quad i=1,2
\;,
\end{equation}
which holds
for example if all particles are assumed to sit in a $1$-dimensional harmonic oscillator (HO)
potential and the zero point energy is neglected.
Usually we will set $\hbar\omega=1$ in what follows, but
later we consider a scaling law $\omega=\omega(N)$ in the context of the
thermodynamic limit $N\rightarrow\infty$.

The interaction between particles of different sorts will have the simple form
\begin{equation}\label{D.2}
V^{(12)} \phi_n^{(1)}\otimes \phi_m^{(2)}
=
-V
 \phi_n^{(1)}\otimes \phi_m^{(2)}
 \delta_{nm}
\;,
\end{equation}
where $V>0$ is a real parameter.
Thus the systems gains the binding energy $V$ if two
particles of different sorts have the same quantum numbers
and form a compound, say, a ``molecule". The molecules can be viewed
as pseudo-particles in a HO potential with double spacing $2\hbar\omega$
of the energy values.

We consider $N$ fermions (resp.~bosons) of each sort, but particles of
different sorts are still assumed to be distinguishable. The total Hilbert space will thus be
\begin{equation}\label{D.3}
\Hil^{\pm} =
\left( \bigotimes_{\nu=1}^{N}{\Hil^{(1)}} \ \right)^{\pm}
\otimes
\left( \bigotimes_{\nu=1}^{N}{\Hil^{(2)}} \ \right)^{\pm}
\;,
\end{equation}
where the $^{\pm}$sign stands for the completely symmetric ($+$), resp.~antisymmetric ($-$)
subspace.
The model is defined by the Hamiltonian
\begin{equation}\label{D.4}
H=\sum_{\nu=1}^{N}{H_{\nu}^{(1)}}
+
\sum_{\mu=1}^{N}{H_{\mu}^{(2)}}
+
\sum_{\mu,\nu=1}^{N}{V_{\mu\nu}^{(12)}}
\;,
\end{equation}
where the subscripts $\mu,\nu$ indicate the factor of
the tensor product where the $1$- and $2$-particle
Hamiltonians of the form (\ref{D.1}) and (\ref{D.2})
act upon. Clearly, (\ref{D.4}) is invariant under
permutations $\pi\in{\mathcal S}_N \times {\mathcal S}_N$
and will be understood to be restricted to the
subspace $\Hil^{\pm}$.

In the fermionic case the eigenstates of $H$ can be chosen as the products of Slater
determinants
\begin{equation}\label{D.5}
\Phi_{\bf n m}=
\bigwedge_{i=1}^{N}{\phi_{n_i}^{(1)}}
\otimes
\bigwedge_{j=1}^{N}{\phi_{m_j}^{(2)}}
\;,
\end{equation}
where all quantum numbers of the same sort are strictly ordered,
$n_1<n_2<\ldots<n_N$ and $m_1<m_2<\ldots<m_N$.
We will write the latter condition in the compact form ${\bf nm}\in I^-$.
Analogously one obtains the eigenfunctions in the bosonic case, but here we have
$n_1\le n_2\le\ldots\le n_N$ and $m_1\le m_2\le\ldots\le m_N$ or,
equivalently, ${\bf nm}\in I^+$ .
The corresponding energy eigenvalues are
\begin{equation}\label{D.6}
E_{\bf nm}=
\sum_{i=1}^{N}{n_i}
+
\sum_{j=1}^{N}{m_j}
-
V \sum_{i,j=1}^{N}{\delta_{n_i m_j}}
\;.
\end{equation}

\section{The partition function\label{Z}}

The above defined model is admittedly a rather crude
description of a real quantum gas, but it allows
an explicit calculation of its partition function without
any further approximation. As usual, the partition function
is defined by
\begin{equation}\label{Z.1}
{\mathcal Z}_N^\pm(\beta)
=
\sum_{{\bf n m}\in I^\pm}{\exp\left( -\beta E_{\bf nm}\right)}
\;.
\end{equation}
In view of (\ref{D.6}) we will split the series (\ref{Z.1})
into a finite number of subseries according to the number
of coincidences between the two sequences of quantum numbers
${\bf n}$ and ${\bf m}$. More precisely,
we will split the total index set $I^\pm$ or, equivalently,
the set of eigenstates of $H$ into disjoint subsets
according to their ``order type".
The  order type of ${\bf nm}$ comprises the linear
ordering of the union of both sequences ${\bf n}$ and ${\bf m}$,
including the coincidences of quantum numbers. For $N=2$
and the fermionic case we have
$10$ different order types, for example
$(n_1<n_2<m_1<m_2)$
or
$(n_1=m_1<m_2<n_2)$.

In general, every order type can be uniquely characterized
by a string $s$ composed of symbols $s_i\in\{0,1,2\}$.
``$s_i=1$ "  means ``a quantum number of sort $1$",
analogously for ``$s_i=2$ ", but ``$s_i=0$ "
means that ``two quantum numbers of different sort coincide".
In the above example,
$(n_1<n_2<m_1<m_2)\leftrightarrow  (1122)$
and
$(n_1=m_1<m_2<n_2)\leftrightarrow  (021)$.

Let $N_i(s)$ denote  the number of occurences
of the symbol ``$i$" in the string $s$ and $|s|$ the length of $s$.
We define  ${\mathcal A}_N $ as the set of all
finite strings $s$ over the alphabeth $\{0,1,2\}$
such that
$N_1(s)=N_2(s)=N-N_0(S)$.
Then the strings $s$ in  ${\mathcal A}_N $ are in $1:1$ correspondence
with the order types, denoted by $T^\pm(s)$, of the subsequences of (\ref{Z.1}).
Hence we may write
\begin{eqnarray}\label{Z.2}
{\mathcal Z}_N^\pm(\beta)
&=&
\sum_{s\in{\mathcal A}_N}
\sum_{{\bf mn}\in T^\pm(s)}
\exp\left( -\beta E_{\bf mn}\right) \\
&=&
\sum_{N_0=0}^{N}
e^{\beta N_0 V}
\sum_{s\in {\mathcal A}_{N,N_0}}
\sum_{{\bf mn}\in T^\pm(s)}
\exp\left(-\beta\left(\sum_{i}n_i+\sum_{j}m_j\right) \right)
\;.
\end{eqnarray}
In the last equation we have used the definition
\begin{equation}\label{Z.3}
{\mathcal A}_{N,N_0}
=
\{s\in{\mathcal A}_N |N_0(s)=N_0 \}
\end{equation}
and that all strings in ${\mathcal A}_{N,N_0}$
lead to the same interaction energy $-N_0 V$.

Note that the series $\sum_{{\bf mn}\in T^\pm(s)}$
is only a $2N-N_0$-fold series because of the $N_0$
coincidences between the quantum numbers.
In the exponent, however, each term in
$\sum_{i}n_i+\sum_{j}m_j$
has to be counted once, irrespective of coincidences.
Hence we may introduce $2N-N_0$ new summation indices
$a_i$ if we account for the double occurences in the exponent
by factors $g_i$. The condition $a_1<a_2<\ldots<a_{|s|}$ will be
written in the compact form ${\bf a}\in J^-$, analogously
$a_1\le a_2\le \ldots\le a_{|s|}\Leftrightarrow {\bf a}\in J^+$.

This yields
\begin{equation}\label{Z.4}
\sum_{{\bf mn}\in T^\pm(s)}
\exp\left(-\beta\left(\sum_{i}n_i+\sum_{j}m_j\right) \right)
=
\sum_{{\bf a}\in J^\pm}
\exp\left(-\beta\sum_{i}g_i(s) a_i \right)
\;,
\end{equation}
where we have used $|s|=2N-N_0$ and the definition
\begin{equation}\label{Z.5}
g_i(s)=
\begin{cases}
1 & s_i=1 \mbox{ or } 2\\
2 & s_i = 0
\end{cases}
\;.
\end{equation}

The remaining evaluation of (\ref{Z.4})  is straight forward.
We introduce as new summation variables the differences
\begin{eqnarray}\label{Z.5a}
d_1 &=& a_1\\
d_n&=&a_n - a_{n-1},\quad n>1
\end{eqnarray}
and use the partial sums
\begin{equation}\label{Z.6}
G_n(s)=\sum_{i=n}^{|s|} g_i(s)
\end{equation}
since
\begin{equation}\label{Z.7}
\sum_n G_n d_n =  \sum_n G_n (a_n-a_{n-1}) = \sum_n (G_n-G_{n+1}) a_n = \sum_n g_n a_n
\;.
\end{equation}
Further let
\begin{equation}\label{Z.7a}
\varepsilon^\pm=
\begin{cases}
0 & \mbox{ for bosons}(+)\\
1 & \mbox{ for fermions}(-)\\
\end{cases}
\end{equation}
Then we obtain

\begin{eqnarray}\label{Z.8}  \nonumber
\sum_{{\bf a}\in J^\pm}
\exp\left(-\beta\left(\sum_{i}g_i(s) a_i\right) \right)
&=&
\sum_{d_1=0}^{\infty}
\sum_{d_2=\varepsilon^\pm}^{\infty}
\ldots
\sum_{d_{|s|}=\varepsilon^\pm}^{\infty}
\exp\left(-\beta\left(\sum_{i}G_i(s) d_i\right) \right)
\\
&=&
\sum_{d_1=0}^{\infty}
e^{-\beta G_1 d_1}
\sum_{d_2=\varepsilon^\pm}^{\infty}
e^{-\beta G_2 d_2}
\ldots
\sum_{d_{|s|}=\varepsilon^\pm}^{\infty}
e^{-\beta G_{|s|} d_{|s|}}
\label{Z.8a}
\\
&=&
\frac{1}{1-e^{-\beta G_1}}
\prod_{n=2}^{|s|}{\frac{ e^{-\varepsilon^\pm\beta G_n}}{1-e^{-\beta G_n}}}
\;.
\label{Z.8b}
\end{eqnarray}

The result can be further simplified by utilizing the fact
that (\ref{Z.8b}) depends only on the $G_n$, i.~e.~on the
positions of the symbol ``$0$" in the string $s$. There are
exactly ${ 2(N-N_0)\choose N-N_0}$  strings in ${\mathcal A}_{N,N_0}$
with the same positions of ``$0$" and hence the same values (\ref{Z.8b}).
We may thus insert the factor  ${ 2(N-N_0)\choose N-N_0}$ and rather
sum over the set ${\mathcal B}_{N,N_0}$ of all strings $s$
over the alphabeth $\{0,1\}$ satisfying $N_0=N_0(s)=N-N_1(s)/2$.
$g_i(s)$ and $G_i(s)$ are defined analogously as before.
Then the final result reads
\begin{equation}\label{Z.9}
{\mathcal Z}_N^\pm(\beta)=
\sum_{N_0=0}^{N}
e^{\beta N_0 V}
{ 2(N-N_0)\choose N-N_0}
\sum_{s\in{\mathcal B}_{N,N_0}}
\frac{1}{1-e^{\beta G_1(s)}}
\prod_{n=2}^{|s|}{\frac{ e^{-\varepsilon^\pm\beta G_n(s)}}{1-e^{-\beta G_n(s)}}}
\;.
\end{equation}

\section{Specific heat and entropy\label{C}}
\begin{figure}
  \includegraphics[width=\columnwidth]{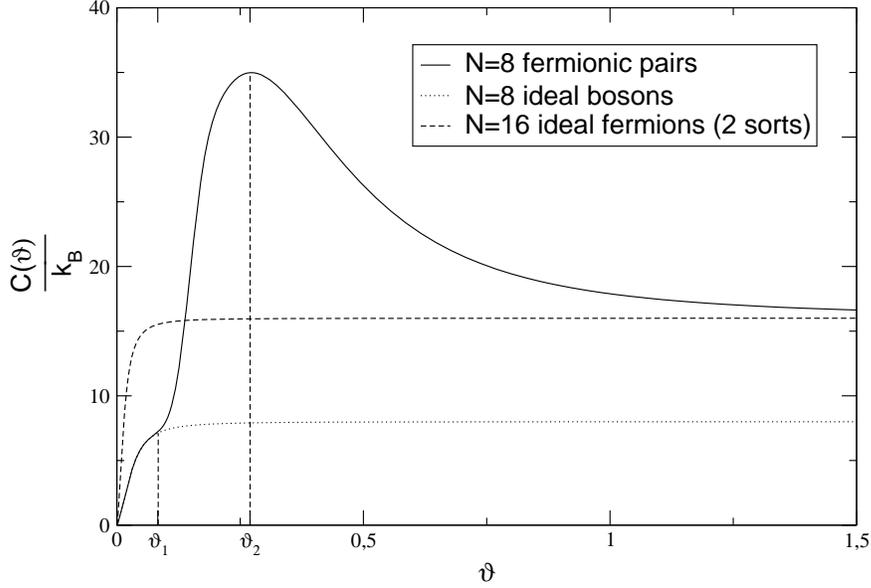}
\caption{\label{fig1} The specific heat $C_8^-$ as a function of
  $\vartheta =\frac{1}{\beta V}$  with $V=100$. This can be compared with the
  specific heat of system of $8$ bosons in a $2\hbar\omega$-HO potential or of
  $16$ fermions in a $\hbar\omega$-HO potential.}
\end{figure}

As usual, we define
\begin{equation}\label{C.1}
C_N^\pm(\beta)
=
\beta^2 \frac{\partial^2}{\partial \beta^2}
\ln {\mathcal Z}_N^\pm(\beta)
\;.
\end{equation}
The explicit form would be too complex to be reproduced here. However,
using a computer algebra software the specific heat
function can be calculated and plotted for small $N$,
say $N\le 8$, without problems. A typical plot of
$C_N^-$  as a function of $\vartheta =\frac{k_B T}{V}=\frac{1}{\beta V}$
is shown in figure \ref{fig1}.

For $\vartheta  \rightarrow 0$ the specific heat vanishes exponentially
as $\exp\left(-\frac{\hbar \omega}{k_B T} \right)$, which is typical for finite quantum systems.
For $\vartheta \approx\vartheta _1$ a first plateau of the height $C_8=8$ is reached which
is a ``preliminary" classical limit of a system of $8$ molecules moving in a HO
potential, in accordance with the law of Dulong-Petit. In the region $0<\vartheta <\vartheta _1$
the molecules are practically never broken into parts and behave as pseudo-particles.
This is confirmed by a comparison with the specific heat function of $8$ bosons (or fermions)
in a $2\hbar\omega$-HO potential, see figure \ref{fig1}. (For the identity
$C_N^{-}=C_N^{+}$ in 1-dimensional HO potentials see \cite{ScS:PA98}.)
For $\vartheta>\vartheta_1$ the specific heat increases above its preliminary classical level
which indicates a beginning dissociation of the molecules. For $\vartheta\approx\vartheta_3$
the final classical limit $C_6=16$ of $16$ particles in a HO potential is reached and all molecules
are dissociated. But before, at $\vartheta\approx\vartheta_2$, a marked peak of the specific heat
occurs which means that at this temperature the maximal portion of any energy supply would be used for
the dissociation of the molecules instead of heating up the system.
This peak at $\vartheta\approx\vartheta_2$ might indicate a diffuse phase transition,
however, the question whether a real phase transition occurs can only be addressed by considering
the thermodynamic limit $N\rightarrow\infty$ in the next section
and will be answered negatively there.\\

\begin{figure}
  \includegraphics[width=\columnwidth]{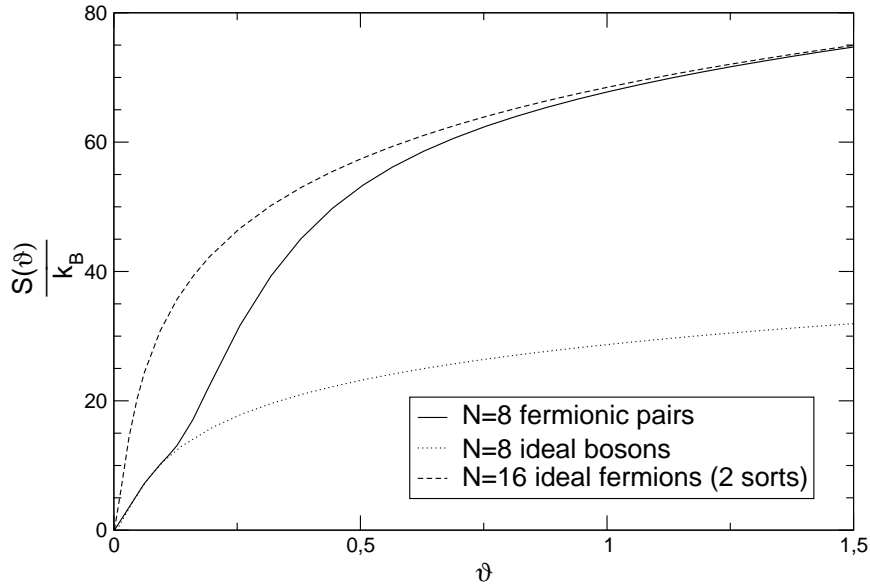}
\caption{\label{fig4} The entropy $S$ as a function of $\vartheta =\frac{1}{\beta V}, V=100,$
for $8$ fermion pairs. Compare the corresponding plots of the entropy of an ideal gas of $16$ fermions
of two different sorts for high temperature and of $8$ bosons for low temperature. }
\end{figure}

The entropy function $S(T)$ interpolates between the entropy
of $8$ ideal bosons for low temperature and of $16$ fermions of two different sorts for high temperature,
see figure \ref{fig4}.

\section{The thermodynamic limit\label{L}}

It turns out that the explicit solution (\ref{Z.9}) can also be evaluated in the limit
$N\rightarrow\infty$. But we have to perform  a simultaneous scaling of the energy
level spacing $\hbar \omega$.
This is analogous to the usual practice of letting the volume of a system go to infinity
in such a way that the particle density remains finite.

To find the correct scaling law in our case, we consider $N$ fermions occupying
the first $N$ eigenstates of a HO potential (the ``Fermi sea").
Their $1$-particle density assumes values which diverge as $\sqrt{N}$ within an interval
of length $\sim \sqrt{N}$ in accordance with the total integral of the $1$-particle density being $N$.
The typical unit of length of the HO is $x_0=\sqrt{\frac{\hbar}{m \omega}}$. Hence a scaling
$\omega \rightarrow \frac{\omega}{N}$  stretches the system by a factor $\sqrt{N}$ and reduces
the $1$-particle density to  finite values for $N\rightarrow\infty$. For a more detailed account
see \cite{GWSZ:00}. In the case of a boson gas we cannot argue in the same way as above, but we will
choose the same scaling. In this case the existence of a thermodaynamic limit, which is shown below,
is the only justification of choosing the scale transformation $\omega \rightarrow \frac{\omega}{N}$ .

In practice the considered scaling means that the variable $\beta$ on the rhs of (\ref{Z.9})
has to be replaced by $\frac{\beta}{N}$, except for the factor $e^{\beta N_0 V}$ which comes from the
interaction, not from the HO potential. Consequently, the corresponding terms can be approximated by
their high temperature limit, i.~e.~by their first order Taylor expansion with respect to $\beta/N$.
In this limit the difference between the statistics of fermions and bosons disappears.

Hence
\begin{equation}\label{L.2}
\frac{e^{-\varepsilon^\pm\frac{\beta}{N}G_n(s)}}
{1-e^{-\frac{\beta}{N}G_n(s)}}\simeq \frac{N}{\beta G_n(s)}
\quad\beta\mbox{ fixed, } N\rightarrow\infty
\end{equation}
and we obtain
\begin{eqnarray}\label{L.3a}
{\mathcal Z}_N(\beta)
&\simeq &
\sum_{N_0=0}^{N}
e^{\beta N_0 V}
{ 2(N-N_0)\choose N-N_0}
\sum_{s\in{\mathcal B}_{N,N_0}}
\prod_{n=1}^{|s|}{\frac{N}{\beta G_n}} \\
\label{L.3b}
&=&
\sum_{N_0=0}^{N}
e^{\beta N_0 V}
{ 2(N-N_0)\choose N-N_0}
\left(\frac{N}{\beta}    \right)^{2N-N_0}
\sum_{s\in{\mathcal B}_{N,N_0}}
\prod_{n=1}^{|s|}{\frac{1}{G_n}}
\;.
\end{eqnarray}

In the next step we calculate the combinatorical factor:
\begin{equation}\label{L.4}
\sum_{s\in{\mathcal B}_{N,N_0}}
\prod_{n=1}^{|s|}{\frac{1}{G_n}}
=
\frac{1}{2^N_0 N_0! (2N-2N_0)!}
\;.
\end{equation}
In order to prove (\ref{L.4}) we first consider an example.
Let $N=6, N_0=3, N_1=6$ and consider the string $s=(101100111)\in{\mathcal B}_{N,N_0}$. We
then have $(g_n)= (121122111)$ and $(G_n)=(12,11,9,8,7,5,3,2,1)$.
Hence $\prod_{n}\frac{1}{G_n}=\frac{10\cdot 6 \cdot 4}{12!}$.
Generalizing this example, we see that $\prod_{n}\frac{1}{G_n}$ can always be written as a fraction
with denominator $(2N)!$ and the nominator being a product of $N_0$ integers $n_\nu$
with $1\le n_\nu < 2N$ such that their distance is $|n_\nu-n_\mu|\ge 2$ for $\nu\neq \mu$.
Conversely, any $N_0$ integers with these properties occur in the nominator of the terms in the sum
$\sum_{s\in{\mathcal B}_{N,N_0}}\prod_{n=1}^{|s|}{\frac{1}{G_n}}$. We set $2N=M$ and generalize to arbitrary
natural numbers $M$. Let $F(M,N_0)$ be defined as the sum of all products of $N_0$-tuples
$(n_1,\ldots,n_{N_0})$ with $n_1<n_2<\ldots<n_{N_0}<M$ and $|n_\nu-n_{\nu+1}|\ge 2$.

If we enlarge $M$ by $1$  additionally those $N_0$-tuples satisfy the above condition with $n_{N_0}=M$.
Hence the following recursion relation holds for $F$:
\begin{equation}\label{L.5}
F(M+1,N_0)=F(M,N_0)+M F(M-1,N_0-1)
\;.
\end{equation}
Moreover,
\begin{equation}\label{L.6}
F(M,N_0)=0 \mbox{ for }M<2 N_0
\end{equation}
and
\begin{equation}\label{L.7}
F(M,0)=1 \mbox{ for all }M
\;.
\end{equation}
The recursion relation (\ref{L.5}) and the initial conditions  (\ref{L.6}), (\ref{L.7})
are satisfied by the following solution
\begin{equation}\label{L.8}
F(M,N_0)=\frac{M!}{2^{N_0}N_0!(M-2N_0)!},
\end{equation}
as can be easily confirmed. Setting $M=2N$ and dividing by $(2N)!$ completes
the proof of (\ref{L.4}).

Inserting (\ref{L.4}) into (\ref{L.3b}) yields

\begin{equation}\label{L.8a}
{\mathcal Z}_N(\beta)
\simeq
\sum_{N_0=0}^{N}
e^{\beta N_0 V}
\left(\frac{N}{\beta}  \right)^{2N-N_0}
\frac{1}{2^N_0 N_0! (N-N_0)!^2}
\;.
\end{equation}

Next we approximate the factorials in (\ref{L.8a}) by Stirling expressions,
e.~g.~$N_0!\approx N_0^{N_0}e^{-N_0}$, set $N_0=Nx$ and replace the sum
$\sum_{N_0=0}^{N}\ldots$ by an integral $\int_0^{1}N\, dx\ldots$. The logarithm
of the integrand $I(x)$ reads
\begin{equation}\label{L.9}
\ln I(x)=
N\left\{
2-x +V\beta x- x\ln 2 +2(x-1)\ln(1-x)-x\ln x -2 \ln \beta+x\ln\beta
\right\}
\;.
\end{equation}
The term in brackets has as a maximum at
\begin{equation}\label{L.9a}
x_0=\frac{1}{\beta} e^{-V\beta}
\left(
1+\beta  e^{V\beta}-\sqrt{1+2\beta  e^{V\beta}}
\right)
\;,
\end{equation}
hence the integrand will be sharply peaked at this maximum if $N\rightarrow\infty$.
Thus we can evaluate the integral for $N\rightarrow \infty$ using the saddle point method.

In the last step we remove all factors in the partition function approximation which are not
of the form $\zeta^{N}$, $\zeta$ being independent of $N$. These factors would only give finite
$N$ corrections in the thermodynamic functions and have to be neglected in the thermodynamic limit.
The final result is
\begin{equation}\label{L.10}
{\mathcal Z}_\infty(\beta)
=
2^{-2N} \exp\left\{
\frac{N}{\beta} e^{-V\beta}
\left(
e^{V\beta}\beta-1+\sqrt{1+2e^{V\beta}\beta}
\right)
\right\}
\left[
\frac{\beta}{1+\sqrt{1+2e^{V\beta}\beta}}
\right]^{-2N}
\;.
\end{equation}

\begin{figure}
  \includegraphics[width=\columnwidth]{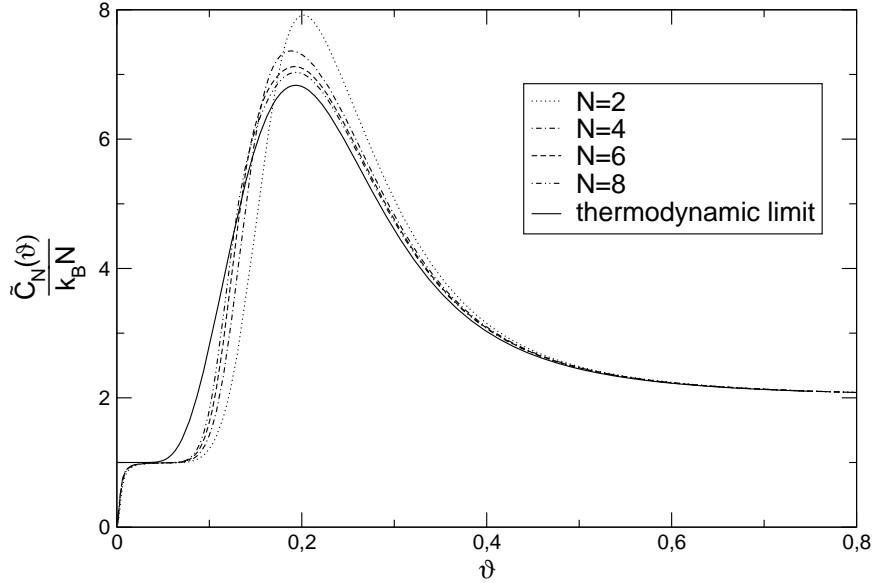}
\caption{\label{fig2} The specific heat per molecule $\widetilde{C}_N^-/N$ as a function of
  $\vartheta =\frac{1}{\beta V}$  with $V=100$ for $N=2,\ldots,8$ which rapidly
  converge to the
  thermodynamic limit function $c(\beta)$, except for low temperatures.
  The tilde in $\widetilde{C}_N^-/N$ indicates
  the scaling $\omega \rightarrow \frac{\omega}{N}$.}
\end{figure}

The resulting specific heat per molecule will be independent of $N$:
\begin{eqnarray}
&c(\beta)&       \label{L.11a}
=
\frac{\beta^2}{N} \frac{\partial^2}{\partial \beta^2}
\ln {\mathcal Z}_\infty(\beta)\\  \label{L.11b}
&=&
\frac{2(2(1+\sqrt{1+2e^{V\beta}\beta})+e^{V\beta}\beta(5+2V\beta+V^2\beta^2+\sqrt{1+2e^{V\beta}\beta}))}
{\sqrt{1+2e^{V\beta}\beta}(1+\sqrt{1+2e^{V\beta}\beta})^2}
\end{eqnarray}
The rapid convergence of the finite $N$ specific heat per molecule, scaled with
$\omega \rightarrow \frac{\omega}{N}$, to the limit (\ref{L.11b}) is
illustrated in figure \ref{fig2}.

The limits
$\lim_{\beta\rightarrow 0}c(\beta)=2$ and
$\lim_{\beta\rightarrow \infty}c(\beta)=1$
are in accordance with our discussion in section \ref{C}.

\begin{figure}
  \includegraphics[width=\columnwidth]{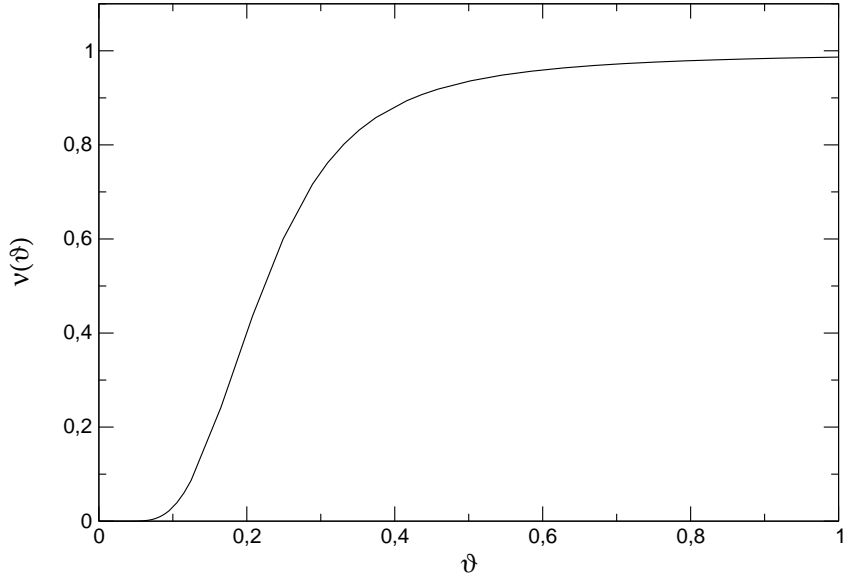}
\caption{\label{fig3} The degree of dissociation $\nu$ as a function of
$\vartheta =\frac{1}{\beta V}, V=100,$
in the thermodynamic limit.}
\end{figure}

We define the ``degree of dissociation" $\nu(\beta)$ as the thermal
expectation value of $1-\frac{N_0}{N}=1-x$, or, equivalently, as
\begin{equation}\label{L.12}
\nu(\beta)=1-\frac{1}{N\beta}\frac{\partial}{\partial V}\ln {\mathcal Z}_\infty(\beta)
\end{equation}
and obtain the simple result
\begin{equation}\label{L.13}
\nu(\beta)=\frac{2}{1+\sqrt{1+2e^{V\beta}\beta}}
\;.
\end{equation}
A plot of the degree of dissociation as a function of $\vartheta$,
see figure \ref{fig3}
confirms our previous interpretation of the specific heat function.

Since  (\ref{L.10})  is a smooth function of $\beta$ there is no phase transition in our model.
Instead of a critical temperature at which the molecule phase is changing into the fermi gas phase
we have an extended temperature domain where the transition occurs, even in the thermodynamic limit.
The reason of this may may be that our model is only $1$-dimensional and, moreover, is not suited to
describe long range order.
\section{Summary and outlook\label{S}}

We have studied a two-species quantum gas of $2N$ particles, fermions or bosons, confined in a $1$-dimensional
HO-potential with a simplified attractive $2$-body interaction between particles of different sorts. The
canonical partition function and hence the relevant thermodynamical functions of this system can be written
in terms of finite sums and products and explicitely calculated by using computer-algebraic means for $N\le 8$.
The specific heat function shows a ``preliminary" classical saturation value of $c\sim N k_B$
for medium temperature, a peak indicating the increasing dissolution of the pairs and a high temperature
limit of $c\sim 2 N k_B$.  Using the scaling
$\omega \rightarrow \frac{\omega}{N}$ we calculated the thermodynamic limit, i.~e.~the limit of
$\frac{1}{N}\ln Z_N^\pm$ and related functions for $N\rightarrow\infty$. In this limit the
difference between bosons and fermions vanishes. The peak of the specific heat function remains
smooth and no phase transition occurs.\\

We motivated the construction of our model by considering the problem of pseudo-bosons, that is,
the problem, in which sense pairs or, more generally,
even numbers of fermions could be considered as bosons.
What can be learnt from our model concerning this question? As expected,
the behavior of the specific heat function mentioned above deviates
from that of an $N$-particle bose gas, see figure \ref{fig1},
similarly for the entropy function, see figure \ref{fig4}.

However, for low temperatures the difference between bosons and pseudo-bosons is not so marked as we would
expect in the general case. This is partly due to the circumstance that the specific heat functions of ideal
fermi and bose gases in $1$-dimensional HO potentials are the same, see \cite{ScS:PA98}.
Only the ground state energies differ by a constant. Also the entropy function $S(T)$ of the pseudo-bose gas
is very well approximated by the corresponding bose gas function at low temperatures.
We expect that the differences would
become more marked in the $3$-dimensional HO potential case, where Bose-Einstein condensation occurs.
Unfortunately, it is not clear how to generalize the method used in this article to that case, and the problem
of pseudo-Bose-Einstein condensation remains unsolved for the present.
\section*{Acknowledgment}
We would like to thank Klaus B\"arwinkel and J\"urgen Schnack for their critical reading of our manuscript.


\end{document}